\documentclass{article}

\usepackage{arxiv}
\makeatletter
\AtBeginDocument{%
  \pagestyle{fancy}%
  \rhead{}%
}

\renewcommand{\keywords}[1]{\par\noindent\textbf{Keywords:} #1}
\makeatother
\usepackage{booktabs}
\usepackage{tabularx}
\usepackage{tikz}
\usepackage{tikz-3dplot}
\usepackage[utf8]{inputenc} 
\usepackage[T1]{fontenc}    
\usepackage{hyperref}       
\usepackage{url}            
\usepackage{booktabs}       
\usepackage{amsfonts}       
\usepackage{nicefrac}       
\usepackage{microtype}      
\usepackage{lipsum}
\usepackage{graphicx}
\usepackage{enumitem}
\usepackage{subcaption}
\usepackage{amsmath}
\usepackage{array} 
\usepackage{natbib}
\usepackage{lineno}         

\usepackage{setspace}
\usepackage{tikz}
\usetikzlibrary{arrows.meta,positioning,calc}
\usepackage{pgfplots}
\pgfplotsset{compat=1.18}
\usepackage{tikz-3dplot}

\usepackage{authblk}

\makeatletter
\renewcommand\AB@affilnote[1]{\textsuperscript{#1}\,}
\makeatother

\usepackage{tikz}
\usetikzlibrary{arrows.meta,positioning,calc,shapes.geometric}
\usepackage{tikz-3dplot}
\usepackage{booktabs}
\usepackage{tabularx}

\usepackage{etoolbox}

\makeatletter
\patchcmd{\@maketitle}{\@date}{}{}{}
\makeatother

\author[1,2,3]{Andrea Ferrario\thanks{Corresponding author: \texttt{aferrario@ethz.ch}.} }
\affil[1]{Institute of Biomedical Ethics and History of Medicine, University of Z\"urich, Z\"urich, Switzerland}
\affil[2]{SUPSI, Dalle Molle Institute for Artificial Intelligence (IDSIA), Lugano, Switzerland}
\affil[3]{ETH Z\"urich, Z\"urich, Switzerland}

\title{High-Risk AI Systems and the Problem of Identity in the European AI Act}

\makeatletter
\renewcommand{\@maketitle}{%
  \vbox{%
    \hsize\textwidth
    \linewidth\hsize
    \vskip 0.1in
    \@toptitlebar
    \centering
    {\LARGE\sc \@title\par}
    \@bottomtitlebar
    \textsc{\undertitle}\\
    \vskip 0.1in
    \def\And{%
      \end{tabular}\hfil\linebreak[0]\hfil%
      \begin{tabular}[t]{c}\bf\rule{\z@}{24\p@}\ignorespaces
    }%
    \def\AND{%
      \end{tabular}\hfil\linebreak[4]\hfil%
      \begin{tabular}[t]{c}\bf\rule{\z@}{24\p@}\ignorespaces
    }%
    \begin{tabular}[t]{c}\bf\rule{\z@}{24\p@}\@author\end{tabular}%
    \vskip 0.2in
  }%
}
\makeatother

\newcommand{\undertitle}{}

\begin{document}

\maketitle

\vspace{-1em}

\begin{center}
Accepted as a non-archival paper at \emph{The 2026 ACM Conference on Fairness, Accountability, and Transparency} \\ (FAccT '26), June 25--28, 2026, Montreal, QC, Canada.
\end{center}

\begin{abstract}
The EU Artificial Intelligence Act (AIA) establishes a lifecycle governance regime for high-risk AI systems built around ex-ante conformity assessment, post-market monitoring, and re-assessment upon ``substantial modification.'' These obligations presuppose AI identity judgments: regulators and providers must decide when an updated system remains the same system over time. In this work, we show how this logic is clarified by the \texttt{function}\textsuperscript{\tiny +} framework of artifact identity, which individuates AI systems by their techno-function together with context-sensitive criteria of appropriate functioning, captured as ``AI trustworthiness.'' We further argue that the AIA does not provide an internal, auditable criterion for \emph{synchronic} identity---when two AI systems at a given time should count as the same for regulatory purposes---and instead largely defers such sameness determinations to sectoral or harmonization instruments. \texttt{function}\textsuperscript{\tiny +} supplies a synchronic identity test anchored in techno-function and trustworthiness profiles and levels, making synchronic identity decisions inspectable in governance settings such as procurement, liability, and market surveillance. Our contribution is a conceptual and auditing lens: we provide a correspondence map between AIA lifecycle obligations and \texttt{function}\textsuperscript{\tiny +} identity components, and we make the synchronic case operationally legible via a minimal decision flow for audit and dispute contexts. We conclude with two implementation-facing recommendations: (1) more precise, testable reporting of intended purpose, and (2) standardized, auditable trustworthiness reporting that supports comparability over time and across deployments.
\end{abstract}

\keywords{Artificial Intelligence, EU AI Act, AI Governance, Metaphysics, Identity, Time, Audit}

\section{Introduction}
\label{section:intro}
The European Union Artificial Intelligence Act (AIA) \citep{EU_AI_Act_2024} is the first regulatory framework for governing AI in the EU market, aiming to safeguard health, safety, and fundamental rights while promoting trustworthy AI (Recital 1, AIA). Notably, the regulation classifies systems into four risk categories while introducing a distinctive regulatory lifecycle regime for the so-called high-risk AI systems. This regime 
hinges on ex-ante conformity assessment, post-market monitoring, and renewed assessment when a change constitutes a ``substantial modification'' (Article~3(20), Annex~IV; Article~3(23) and Recital~128, AIA). In practice, this regime requires repeated judgments about whether an AI system is still ``the same'' system for regulatory purposes as it is updated, re-trained, or re-configured.

We refer to these as ``AI identity'' judgments in two senses. \emph{Diachronic identity} concerns persistence: when does an updated AI system remain the same system over time, as opposed to becoming (legally) a new system that must undergo a new conformity assessment? \emph{Synchronic identity} concerns sameness-at-a-time: when should two contemporaneous instantiations (e.g., copies, configurations, or provider variants) be treated as the same AI system for regulatory comparability? The latter question becomes unavoidable whenever governance actors must decide whether compliance evidence, documentation, monitoring signals, or corrective actions established for one instantiation legitimately generalize to another---for example across certification scoping, traceability and incident response, or comparability claims in procurement and market surveillance.

Motivated by the AIA's regulatory approach to AI identity over time, this paper advances two claims and is explicitly scoped as a conceptual and auditing contribution. We do not propose a validated compliance procedure or an empirical implementation study. Rather, we aim to make explicit the identity assumptions that are already doing work in AIA lifecycle governance and to provide a structured lens for rendering identity-relevant determinations more transparent and contestable. \textbf{Claim 1} is that the AIA's lifecycle mechanism implicitly operationalizes diachronic identity for high-risk AI systems. Conformity assessment fixes an identity-relevant baseline (intended purpose and requirements of appropriate functioning), post-market monitoring tracks continued compliance, and ``substantial modification'' functions as a legal discontinuity condition that triggers re-certification. We show how this structure is clarified by the \texttt{function}\textsuperscript{\tiny +} framework of artifact identity \citep{carrara2009fine,ferrario2025trustworthiness}, which individuates AI systems by their techno-function together with context-sensitive criteria of appropriate functioning (a trustworthiness profile) and their satisfaction over time (a trustworthiness level). Figure~\ref{fig:AIA_reg} summarizes this correspondence from AIA lifecycle obligations, through governance-to-metaphysics identity questions, to \texttt{function}\textsuperscript{\tiny +} identity components. \textbf{Claim 2} is that the AIA does not provide an internal, auditable criterion for synchronic identity for stand-alone high-risk systems and instead largely defers sameness determinations to sectoral or harmonization instruments. This allocation is often reasonable, but it can leave under-specification precisely where governance must adjudicate sameness claims across contemporaneous instantiations. Figure~\ref{fig:claim2-flow} makes this problem operationally legible by depicting a minimal audit workflow: sameness claims must be justified at conformity assessment (purpose/profile) and maintained under post-market monitoring (trustworthiness level at time $t$), with explicit ``no'' points that terminate comparability.

Our contributions are therefore: (1) a legally grounded analysis of how identity assumptions are embedded in AIA lifecycle governance; (2) a structured conceptual alignment between the AIA mechanism and \texttt{function}\textsuperscript{\tiny +} that renders identity-relevant determinations more transparent and auditable; and (3) two implementation-facing recommendations focused on documentation and comparability: more precise, testable reporting of intended purpose (including for systems with generative capabilities), and standardized, auditable trustworthiness reporting that supports comparability across deployments and over time. While an increasing body of literature focuses on ethical challenges and legal shortcuts affecting the AIA \citep{floridi2021european,smuha2021eu,laux2024trustworthy,worsdorfer2025eu,loi2025regulating}, our paper provides metaphysical grounding for the AIA's identity-relevant lifecycle logic and frames synchronic identity in a way that is inspectable in audit and dispute settings.

The remainder of the paper proceeds as follows. Section~\ref{section:motivating_examples} introduces two motivating cases. Section~\ref{section:Meta_TW} presents \texttt{function}\textsuperscript{\tiny +} and the trustworthiness-based characterization of AI systems. Sections~\ref{section:AIA_intro}-\ref{section:AIA_Identity_Diachronic} summarize the relevant AIA provisions. Sections~\ref{section:claim_1}-\ref{section:claim_2} develop Claims~1-2. Section~\ref{section:recommendations} provides recommendations, and Section~\ref{section:conclusions} shares our conclusions.

\section{Motivation:\ Two Illustrative Cases of AI System Identity Problems}
\label{section:motivating_examples}
To motivate our analysis of AI identity, we present two stylized but realistic cases involving high-risk AI systems (here, systems used in sensitive applications; the term is defined below). The cases are not intended as reports of actual disputes, but as simplified scenarios that isolate identity questions. The first concerns identity over time (diachronic identity); the second, identity at a given time (synchronic identity).

\vspace{1em}

\noindent\textbf{1.\ Changes over time in AI systems for clinical triage}.\ Consider an AI system developed by a company and deployed in the emergency department of a hospital in an EU country.\ The system outputs a binary classification of incoming patients into priority levels: low and high. (This strong simplification suffices for the present case.) The classifications feed into a triage process which seeks to maximize life years and optimize emergency resource allocation.\ At deployment, the system exhibits stable performance, as documented by the provider during the extensive pre-market testing that granted the system official certification.

However, a subsequent respiratory disease outbreak disproportionately affects older patients and causes a distributional shift: the real-world patient data now diverge from the distributions used to develop the system.\ In response, the provider implements a series of overnight updates, retraining the core model on new hospital data for about a month.\ Furthermore, different fairness constraints are introduced to mitigate age-related disparities and the explainability module  includes counterfactual explanations.\footnote{A counterfactual explanation of why $Y$ got a low priority score and, therefore, was denied resource-intensive emergency treatments could look like this: ``if patient $Y$ were ten years older and their illness-severity score below four, then they would have received a high priority score,  other things equal.''}  In parallel, hospital data engineers apply further modifications through a dedicated API, outside the direct oversight of the provider.\ Over time, a chart review reveals anomalous phenomena.\ Clinicians observe that the counterfactual explanations and model outcomes of different patients now tautologically align, undermining their role as actionable explanations in discussions with patients' family members.\ Although the provider insists that the system remains ``the same,'' and its accuracy remains stable over time, many clinicians believe it is  substantially different from the original version and argue that it requires a full, and costly, re-assessment.\ Some simply state that they do not trust the system anymore.\ The issue eventually escalates to hospital management, while patient advocacy groups and journalists begin investigating the case: the provider has sold licenses for the AI system to twelve healthcare institutions in the same country.

At what point does an AI system cease to be the ``same'' system and instead count as a \emph{substantially} and \emph{legally} ``new'' one? Do changes to intended purpose, model architecture, or explainability modules alter identity, and if so, how?  Third-party liability, regulatory compliance, and insurance coverage all hinge on how this distinction is drawn.\ It seems that criteria relating system design requirements to its post-deployment monitoring are needed to govern such phenomena.

\vspace{1em}

\noindent\textbf{2.\ When are AI systems used in biometric border control the same?}
Consider a public tender for real-time biometric identification systems in major European airports.\ National authorities are willing to procure two high-risk AI systems---let us call them $A$ and $B$---from the same vendor for two different airports. While revising the system documentation provided by the vendor, the procurement officials notice that both systems rely on the same open-source models for generating facial bounding boxes and classifying images within those boxes.\ The main differences lie in the user interface and some minor workflow integrations.\ Importantly, however, system $A$ shows superior performance: due to optimization work at the design stage, it achieves lower inference latency and faster outcome retrieval, which the vendor would use to justify an exceptionally higher monthly licensing fee than for system $B$.\ This said, both systems underwent the same conformity procedures, with identical testing protocols for accuracy, robustness, safety, and reliability.\ Authorities conclude that both systems are, in substance, the same biometric control device: accuracy is the same, and differences in processing speed are not material in airport settings where passengers are queued either for check-in or customs declarations.\ In their interpretation of ``intended purpose'' of the AI system, its operational context and conditions of use carry equal weight with its techno-function, namely, to biometrically identify subjects.\ 
On this reading, authorities issue an equivalence determination: $A$ and $B$ are the same for the intended use, so $A$'s premium  brings no material benefit. As a result, the authority can award to $B$, or require the provider to offer $A$ at the non-premium rate.\ The provider disputes this view, arguing that the optimizations in $A$ justify treating it as a \emph{substantially different} product, as these procedures yield, overall, a higher level of what they termed a ``system operational adequacy'' index in their documentation.\ In their view, this index summarizes the operational efficiency of the system along different dimensions, including throughput and latency; its being higher for $A$ justifies a higher licensing fee for that system.\ In this example, what makes $A$ and $B$ the \emph{same} system for biometric border control? Do specific system functionality dimensions suffice to declare their sameness (or lack of) \emph{substantially} and \emph{legally}? Here, regulators could resolve the dispute by clarifying whether identical intended use and core functionality are sufficient to establish system identity in legal terms, and whether quantitative differences in ``system operational adequacy'' warrant treating them as legally distinct systems.\ Importantly, for contracting authorities, procurement processes risk distortion if bids rest on systems that are not substantially different: it would make it difficult to justify the acquisition of system $A$ instead of $B$ using taxpayer money.\ This could result in a diminished public trust in the responsible institutions, although, in practical terms, for passengers, being screened by system $A$ or $B$ makes no difference. 

\vspace{1em}

These two cases illustrate how questions of AI identity may arise and affect different societal actors. Both concern, albeit in different ways, what counts as the adequate functioning of an AI system.\ Addressing these questions requires precise criteria that can specify conditions of sameness to guide AI regulations.\ Introducing these criteria is possible at the price of engaging with a contested notion in contemporary AI philosophy: ``trustworthiness.''  This is what we will do in the next section, showing a way to address these AI identity problems.

\section{A Trustworthiness-Based Metaphysics of AI Systems}
\label{section:Meta_TW}
\subsection{The traditional perspective on artifactual metaphysics}
\label{subsection:artifact_traditional}
Artifacts---human-made objects designed to serve purposes---have often been denied a metaphysical standing on a par with natural entities such as rocks, animals, and persons \citep{lowe2014how}.\ Two reasons are typically invoked: artifacts admit multiple (intended and unintended) purposes, and the mapping between their purposes and material realizations is many-to-many. Their disassembly and reassembly also fuel classic puzzles of identity, most famously the Ship of Theseus \citep{gallois2016metaphysics}.\ As a result, for a long time philosophers have defended the idea that it makes no sense (metaphysically) to ask ourselves whether \textbf{identity criteria for artifacts}, namely, formal rules to define the persistence of artifacts over time or determine whether two artifacts are the same (artifact), can be appropriately defined.\footnote{These rules are typically written using propositional logic formalism.\ They can be introduced, for instance,  for mathematical entities, such as sets. The Axiom of Extensionality is such an example---see \citep{pinter2014book}. Identity criteria for human beings are an object of investigation in the philosophy of mind domain instead \citep{olson2024personal}.} 
The relation between genuine metaphysics and identity (criteria) is exemplified by Quine's famous quote: ``no entity without identity'' \citep[p.~44]{quine1969ontological}.

\subsection{The \texttt{function}\textsuperscript{\tiny +} approach to artifactual identity}
\label{subsection:function+}
Carrara and Vermaas have challenged this longstanding perspective on the metaphysics of artifacts introducing the \texttt{function}\textsuperscript{\tiny +} framework as a way to account for the dual nature of artifactual identity \citep{carrara2009fine}.\ According to this view, artifact identity depends on two interrelated dimensions.\ The first is the \emph{techno-function}, namely, what the artifact is designed to do, e.g., to fly for airplanes.\ Only artifacts designed for the same techno-function can be compared through identity criteria, as they belong to the same artifactual ``kind'' (e.g., ``airplanes,'' ``robotic surgery systems'').\footnote{Techno-functions can be specified at different levels of detail: broader specifications pick out wider artifact kinds, whereas finer ones carve out narrower ones.} The second is the \emph{contextual embedding}---the ``{\tiny +}'' in \texttt{function}\textsuperscript{\tiny +}---which refers to the social and normative settings in which the artifact is designed and deployed \citep{carrara2009fine}.\ Artifacts may share the same techno-function but differ in identity if they operate in distinct social contexts with different expectations, norms, and evaluative standards.\ This contextual embedding comprises an operational principle (the rules by which components cooperate to realize that function) and a normal configuration (the canonical arrangement judged to embody that principle, typically shaped by design practice and standards).\ Compared with mereological essentialism \citep{wiggins2001sameness}, where two artifacts are equal if and only if they are constituted by the same parts, \texttt{function}\textsuperscript{\tiny +} is more permissive: two tokens can count as the same (artifact) within a kind despite differing materials, provided the function-operational principle-configuration triple aligns \citep{carrara2009fine}.\  That said, Carrara and Vermaas' \texttt{function}\textsuperscript{\tiny +} framework is not the only possible account of artifact identity. Its relevance lies in the specific problem it addresses. Their proposal is developed against an anti-realist challenge to artifact identity: if artifacts are individuated by their functions, then artifact kinds appear too coarse-grained to support genuine identity criteria. This is the worry associated with Wiggins: very different objects can perform the same function, and therefore function alone cannot determine what kind of artifact something is, or whether it persists as the same artifact over time \citep{wiggins2001sameness}. Baker and Elder respond to this anti-realist tendency by defending the metaphysical reality of artifacts and by treating artifact functions as identity-relevant \citep{baker2004ontology,elder2004real}. Carrara and Vermaas accept this realist orientation, but refine it. Function must be supplemented by further features that specify how the function is realized in practice. This is the role of operational principles and normal configurations. The point of the ``{\tiny +}'' is therefore to avoid both material reductionism and function-only coarseness: artifact identity is not fixed by material constitution alone, nor by function alone, but by function together with the relevant structure of realization. Thus, according to Carrara and Vermaas, artifacts enjoy  genuine
metaphysical standing, and identity claims
about them are possible.\ Framed this way, artifact identity links functional aims to their evolving design.

\subsection{ \texttt{function}\textsuperscript{\tiny +} for AI systems}
\label{subsection:function+_AI}
Carrara and Vermaas' ontological  strategy can be transferred to AI systems once their designed function, context of deployment, and standards of appropriate performance are made explicit. On the one hand, the identity of AI systems cannot be fixed by material components alone: models, datasets, interfaces, deployment environments, documentation, and monitoring practices may change while the system remains functionally continuous. On the other hand, function alone is too coarse-grained for regulatory purposes. Many AI systems may share a nominal function while differing in the constraints, safeguards, performance expectations, and operational conditions under which they count as appropriately functioning. The \texttt{function}\textsuperscript{\tiny +} framework can preserve the centrality of techno-function while requiring further criteria that specify how that function is realized and assessed in practice. Motivated by these considerations, Ferrario recently adapted Carrara and Vermaas' \texttt{function}\textsuperscript{\tiny +} framework to AI systems, introducing explicit criteria to govern the identity of these systems.\ His approach hinges on the observation that `AI trustworthiness' encodes the contextual embedding of AI systems, i.e.,  the social and normative settings in which they are designed and deployed.\ This stance on AI trustworthiness has been recently popularized in the human-AI interaction domain \citep{jacovi2021formalizing} and descends conceptually from Hawley's ``trust as commitment'' account \citep{hawley2014trust}.\ As it will become clearer in Section \ref{section:AI_TW_AIA}, it is widely endorsed by AI regulations as well.\ According to Hawley, to ``trust someone to do something is to believe that she has a commitment to doing it, and to rely upon her to meet that commitment'' \citep[p.~10]{hawley2014trust}.\ Commitments typically give us obligations, but they are not necessarily morally-laden.\ In this framing, an AI system is considered worthy of our trust if it upholds a set of commitments contextualized by the system's intended functioning.\footnote{This approach mirrors how we judge  trustworthiness in interpersonal relations: an individual is worthy of trust when he upholds socially recognized commitments, e.g., being an expert in a doxastic domain or upholding moral standards.\ To increase one's trustworthiness in a social context, one typically acts on the relevant capabilities that are seen as trust-enabling.\ 
}
The trustworthiness of an AI system is encoded in its \emph{trustworthiness profile}, a structured set of formal and informal requirements that fix when an AI system counts as appropriately functioning in its intended use and context of applicability \citep{ferrario2025trustworthiness}.\ These profiles span commitments, such as accuracy and robustness over time, safeguards for fairness and bias mitigation, transparency and explainability, auditability and others.\ (These requirements will return in Section \ref{section:AI_TW_AIA} when we introduce the approach to AI trustworthiness in the AIA.) Each commitment is made explicit in appropriate documentation, called ``contract'' following Hawley's terminology, that describes the way and extent to which the AI system would satisfy the commitment over time.\ Finally, the trustworthiness profile of an AI system is quantified by a \emph{trustworthiness level function} that summarizes the evolution over time of the degree of trustworthiness of the system.\ This function operationalizes context: it aggregates diverse---and many, especially for complex systems in high-stakes applications---AI trustworthiness requirements into one signal that supports, in particular, the monitoring of system trustworthiness.\ While some requirements may decrease over time, e.g., a minor drop in accuracy, the trustworthiness level function provides an overarching, quantitative perspective to what extent the AI system is maintaining its correct functioning---see Section 5.1 and Figure 1 in  
\citep{ferrario2025trustworthiness} for a few examples of trustworthiness level functions.

In summary, in Ferrario's formulation of Carrara and Vermaas' \texttt{function}\textsuperscript{\tiny +}  account, AI systems are metaphysically characterized by their (1) techno-function, (2) trustworthiness profile, and (3) trustworthiness level function.\ 
Ferrario's account does not assume that AI systems are bare individuals that can be identified independently of the kind to which they belong \citep{lowe2014how}. Rather, individuals are tokens or instantiations of a kind, and the kind is fixed by the relevant identity criteria. These criteria are governance-based: AI systems belong to the same kind when they share the same techno-function and the same trustworthiness profile. This level of individuation is higher than ordinary product categories such as `chatbots,'' `automated vehicles,'' or ``clinical triage systems.'' Once this kind is fixed, the relevant AI system tokens are the instantiations that fall under it, and diachronic or synchronic identity judgments can then ask whether they \emph{also} preserve the relevant trustworthiness level.
Within this framework, two identity criteria emerge---see definition 5.1 in \citep{ferrario2025trustworthiness}: 

\begin{itemize}
    \item \textbf{Diachronic identity} ($\textbf{DI}_{\texttt{function}\textsuperscript{\tiny +}}$):\ An AI system persists at two times \( t_1 \) and \( t_2 \) if it maintains the same techno-function and trustworthiness profile at \( t_1 \) and \( t_2 \), and the same trustworthiness levels at those times.
    \item \textbf{Synchronic identity} ($\textbf{SI}_{\texttt{function}\textsuperscript{\tiny +}}$):\ Two AI systems  are identical at time \( t \) if they perform the same techno-function, have the same trustworthiness profile and trustworthiness level at $t$.
\end{itemize}

Applying $\textbf{DI}_{\texttt{function}\textsuperscript{\tiny +}}$ and $\textbf{SI}_{\texttt{function}\textsuperscript{\tiny +}}$ to the two use cases from Section \ref{section:motivating_examples} yields: in triage, adding fairness constraints and counterfactual modules alters the trustworthiness profile, while overnight retraining likely reflects a drop in the trustworthiness level.\ Unless these changes were pre-determined in the trustworthiness contracts and the trustworthiness-level function is robust to the observed accuracy drop, $\textbf{DI}_{\texttt{function}\textsuperscript{\tiny +}}$ indicates that the system stops persisting.\ 
In the second use case, $A$ and $B$ can be identical if operational efficiency measured by latency is a trustworthiness contract for both systems.\ In that case, if the trustworthiness levels at the time of assessment---i.e., the levels of ``system operational adequacy'' in the provider's terminology---are equal, then the systems $A$ and $B$ are identical as border control AI systems, following $\textbf{SI}_{\texttt{function}\textsuperscript{\tiny +}}$.

Now that we clarified the metaphysical perspective on AI identity, we explore how AI regulation manages it.\ To do so, we move forward by briefly introducing the AIA, and discussing its management of AI trustworthiness in relation to high-risk AI systems.

\section{A Minimal Introduction to the AIA}
\label{section:AIA_intro}
The AIA establishes a unified legal framework for the development, deployment, and use of AI systems within the EU (Recital 1, AIA).\ Its enforcement aims to promote the uptake of human-centered and trustworthy AI while ensuring the protection of health, safety, fundamental rights and mitigating the potential harms posed by AI systems (Recital 1, AIA).\ The regulation promotes rules that should be in line with the Union's trade commitments, taking into account other regulatory  approaches, such as the European Declaration on Digital Rights and Principles for the Digital Decade \citep{EU2022DeclarationDigitalRights} and the Ethics guidelines for trustworthy AI of the High-Level Expert Group on Artificial Intelligence (AI HLEG) \citep{EU_HLEG_TrustworthyAI_2019}.\ It  applies to providers of AI systems in the EU, irrespective of whether they are established in the Union, so long as their AI systems impact the EU market (AIA, Article 2).\ Thus, the AIA is \emph{aterritorial} \citep{floridi2021european}, fostering the ``Brussels effect,'' namely, the \emph{de facto} although not \emph{de jure} adoption of EU regulatory standards by non-EU countries to streamline cross‑border economic exchanges \citep{Bradford2020BrusselsEffect,Petit2020BigTech}.\ Within the AIA, an \textbf{AI system} is 

\begin{quote}
a machine-based system that is designed to operate with varying levels of autonomy and that may exhibit adaptiveness after deployment, and that, for explicit or implicit objectives, infers, from the input it receives, how to generate outputs such as predictions, content, recommendations, or decisions that can influence physical or virtual environments (Art.~3, \citep{EU_AI_Act_2024}).\footnote{Although this definition has received some criticism, as some argue that it is too broad \citep{kazim2023proposed,mokander2022conformity}, nevertheless, it allows accommodating a wide range of tools, including those with generative capabilities, that are already present or anticipated to enter the EU market.\ The definition includes ``general-purpose AI systems,'' which are defined in Article 3(66)-(69), AIA.}
\end{quote}

Notably, the AIA follows a risk-based approach to the regulation of AI systems in the EU market, whereby the legal obligations of responsible actors are determined by the assigned risk category of each system.\ It distinguishes four categories: low or minimal risk, limited risk, high risk, and unacceptable risk.\footnote{Minimal risk AI systems are not subject to regulatory obligations, while limited risk AI systems must meet general transparency requirements.\ Unacceptable risk AI systems are prohibited from entering the EU market.} \textbf{High-risk AI systems} fall into two categories.\ First, they may either constitute a safety component of a product\footnote{In the AIA's Article 3(14), one reads: ```safety component' means a component of a product or of an AI system which fulfils a safety function for that product or AI system, or the failure or malfunctioning of which endangers the health and safety of persons or property.''} or be themselves  products governed by one of the Union harmonization legislations listed in Annex I of the AIA.\ These legislations cover sectors such as machinery, toys, lifts, medical devices, and civil aviation technology.\ In this case, AI systems inherit their high-risk status from being the safety component of a \emph{regulated} product category (or the product itself). Second, systems are deemed high-risk if their \textbf{intended purpose}, i.e., ``the use for which an AI system is intended by the provider, including the specific context and conditions of use, as specified in the information supplied by the provider [...] in the technical documentation'' (Article 3(12), AIA), are one of the specific cases enumerated in Annex III of the AIA.\ These use cases include biometric identification and categorization of individuals, the management of critical infrastructure, education and vocational training, access and enjoyment of private and public essential services, law enforcement, migration, asylum and border control management, as well as the administration of justice and democratic processes.\ 
These high-risk AI systems are called ``stand-alone.''\footnote{See, for instance, the AIA's Recital 52: ``As regards stand-alone AI systems, namely high-risk AI systems other than those that are safety components of products, or that are themselves products, it is appropriate to classify them as high-risk if, in light of their intended purpose, they pose a high risk of harm to the health and safety or the fundamental rights of persons, taking into account both the severity of the possible harm and its probability of occurrence [...].''}
Over time, the list of stand-alone high-risk AI systems may be updated by the European Commission under the procedure outlined in Article 7 of the AIA.

\section{AI Trustworthiness in the AIA}
\label{section:AI_TW_AIA}
The AI HLEG guidelines on Trustworthy AI form the backbone of the AIA's approach to AI trustworthiness \citep{floridi2021european, smuha2021eu,ho2024eu, worsdorfer2025eu,laux2024trustworthy}.\footnote{The interested reader may consult the literature for an analysis of ``Legally Trustworthy AI'' \citep{smuha2021eu}, the socio-epistemological perspective on trust and trustworthiness promoted in the AIA \citep{ho2024eu}, and the supposed conflation between AI trustworthiness and acceptance of risk in the AIA \citep{laux2024trustworthy}.} These guidelines rest on the non-trivial assumption that AI systems can themselves be objects of human trust, i.e., the same assumption endorsed by \texttt{function}\textsuperscript{\tiny +}, see Section \ref{section:Meta_TW}. They operationalize
AI trustworthiness in four steps. 
First, Trustworthy AI is defined as a \emph{societal standard} that requires systems to be lawful, ethical, and robust throughout their lifecycle.\ Second, this standard is articulated through four \emph{ethical principles}, namely, respect for human autonomy, prevention of harm, fairness, and explicability, alongside a discussion of tensions among them.\ Third, these principles are translated into \emph{seven requirements} of AI trustworthiness: human agency and oversight; technical robustness and safety; privacy and data governance; transparency; diversity, non‑discrimination, and fairness; societal and environmental well‑being; and accountability.\ Finally, the guidelines link these requirements to \emph{technical} and \emph{non-technical methods} for assessing and fostering AI trustworthiness, which have since inspired a broad landscape of frameworks and tools \citep{mattioli2024overview, kaur2022trustworthy, ala2020assessment, zicari2021z, li2023trustworthy, mokander2021ethics,loi2019towards}.\ In particular, the AI HLEG’s Assessment List for Trustworthy AI (ALTAI) operationalizes these requirements as a voluntary self‑assessment tool \citep{ala2020assessment}.\ The AIA refers to the AI HLEG guidelines explicitly in its Preamble.\ For instance, in Recital 7, it is stated that the AIA ``should take into account'' these guidelines, while in Recital 27 the seven requirements are discussed while offering an invitation to operationalize them.\ Finally, Recital 165 encourages providers of high‑risk AI systems to voluntarily adopt additional requirements inspired by the AI HLEG guidelines, among others. In summary, the AIA does not treat trustworthiness as a governance ideal. In fact, trustworthiness is translated into auditable criteria of appropriate functioning. This is where the AIA connects to the \texttt{function}\textsuperscript{\tiny +} framework. The next section and Figure~\ref{fig:AIA_reg} show how this structure is operationalized diachronically through conformity assessment, post-market monitoring, and substantial modification.


\section{Diachronic Identity in the AIA}
\label{section:AIA_Identity_Diachronic}
The AIA regulates the change over time of high-risk AI systems with a governance mechanism that comprises (1) an ex-ante conformity assessment, (2) a post-market monitoring system, and (3) the assessment of ``substantial modifications.'' We discuss these procedures and their relations in the following sections, emphasizing their relation with the AI HLEG guidelines and their Trustworthy AI paradigm.\

\subsection{The ex-ante conformity assessment}
\label{subsection:AIA_conformity_assessment}
The AIA mandates an \textbf{ex-ante conformity assessment} for high-risk AI systems as a prerequisite for their entry into the EU market.\ This assessment is defined as ``the process of demonstrating whether the requirements set out in Chapter III, Section 2 relating to a high-risk AI system have been fulfilled'' (Article 3(20), AIA).\ These requirements are described in some detail in Articles 8-15; they include the design of a risk management system and data governance procedures, the drafting of technical documentation,\footnote{Article 11 reads ``It shall contain, at a minimum, the elements set out in Annex IV.'' We describe the documentation in Annex IV in what follows.} the recording of events over the lifetime of the system, the enforcement of transparency criteria and the provision of information about deployers, human oversight processes, accuracy, robustness and cybersecurity procedures.\ They implement the AI HLEG guidelines' \emph{seven requirements} of Trustworthy AI among others.\
Crucially, if the conformity assessment presented by a provider certifies that the high-risk AI system satisfies these requirements, the system is granted a CE marking as a formal declaration of conformity (Article 44, AIA).\ 
For high-risk AI that is part of a regulated product (or is itself such a product), the conformity assessment can be done under the relevant EU product legislation.\ In contrast, stand-alone high-risk AI systems (Annex III AIA) must follow a separate route: either an internal control procedure or an independent assessment by a third-party notified body---see Figure 2 in \citet{mokander2022conformity}.\footnote{ The internal audit route is available only to stand-alone systems that fully comply with the requirements set out in Chapter 2 of Title III of the AIA \citep{mokander2022conformity}.}

Further, the AIA specifies in Annex IV the technical documentation that is required for the conformity assessment of high-risk AI systems.\ This documentation is rather comprehensive, spanning nine sections.\ Among other obligations, providers must submit a general description of the AI system, including its intended purpose, key components, and the development process (Annex IV, Sections 1–2, AIA).\ 
It also requires providers to document detailed information on the monitoring, functioning, and control mechanisms of the AI system, along with a description of the performance metrics used and their justification (Annex IV, Sections 3-4, AIA).\ These requirements cover multiple AI HLEG guidelines' \emph{technical} and \emph{non-technical methods}.\ Additionally, it mandates disclosure of the risk management system and any relevant changes occurring throughout the system's lifecycle (Annex IV, Sections 5-6, AIA)---an issue we will examine in detail in the following section.\

\subsection{The post-market monitoring system}
\label{subsection:AIA_post_market_monitoring}
Annex IV further specifies that providers must include a list of any applicable harmonized standards, a copy of the EU declaration of conformity, and the system's post-market monitoring plan (Annex IV, Sections 7-8, AIA).\ Central to the monitoring plan is the \textbf{post-market monitoring system}, which is intended to evaluate the performance of high-risk AI systems after they have entered the market (Annex IV, Section 9, AIA).\footnote{For further discussion on the interplay between ex-ante conformity assessment and post-market monitoring, see \citet[Section 4]{mokander2022conformity}.} This monitoring system is designed to continuously collect, document, and analyze data and assess system performance throughout its operational lifetime, drawing on information provided by deployers and other relevant sources.\ Its goal is to allow providers to evaluate ``the continuous compliance of AI systems with the requirements set out in Chapter III, Section 2'' (Article 72(2), AIA).\ 
However, at the time of writing, not much else is known about the monitoring plan and system as only their scope and general governance recommendations are provided.\ That said, the EU Commission plans to establish a template for the post-market monitoring plan in the course of 2026.

\subsection{The `substantial modifications' of high-risk AI systems}
\label{subsection:sub_mod}
The AIA characterizes the possible changes---referred to as ``modifications'' therein---affecting CE-certified high-risk AI systems throughout their lifetime by using the concept of \textbf{substantial modification}, which is  defined in Article 3(23) as follows  (emphasis ours):

\begin{quote}
`substantial modification' is a change to an AI system after its placing on the market or putting into service which is \emph{not foreseen or planned in the initial conformity assessment} carried out by the provider and as a result of which the compliance of the AI system with the requirements set out in Chapter III, Section 2 
is affected or results in a modification to the \emph{intended purpose} for which the AI system has been assessed.
\end{quote}

This concept is the AI-specific translation of substantial modification already regulated in Union harmonization legislation. 
For instance, under the Machinery Regulation \citep{EU2023Machinery}, a substantial modification refers to a change to machinery\footnote{Machinery is defined as an assembly of linked parts that satisfies specific functional relations with energy sources, its working environment, and with software components—see Article 3(1) in \citep{EU2023Machinery}.} that is not foreseen or planned by the manufacturer and that affects the safety of the machinery by creating a new hazard or increasing an existing risk (Article 3). Similarly, under the Cyber Resilience Act \citep{EU2024CyberResilience}, a substantial modification to a product with digital elements is defined as a change that either affects the product's compliance with the cybersecurity requirements listed in that regulation, or alters its intended purpose.\ 
The key point is that substantial modifications of high-risk AI systems result in the emergence of a ``new'' AI system that requires a ``new'' conformity assessment. In fact, Recital 128 reads (emphasis ours):

\begin{quote}
In line with the commonly established notion of substantial modification for products regulated by Union harmonization legislation, it is appropriate that whenever \emph{a change occurs which may affect the compliance of a high-risk AI system with this Regulation} (e.g. change of operating system or software architecture), or when \emph{the intended purpose of the system changes}, that AI system should be considered to be a new AI system which should undergo a new conformity assessment. 
\end{quote}

Thus, substantial modifications force providers of high-risk AI systems to finalize new conformity assessments and wait for new CE certificates.\ Recital 128 in the AIA also describes the changes that \emph{do not} interrupt persistence of a high-risk AI system and, therefore, do not require a new conformity assessment: these are all modifications to the system that have been  pre-determined by the provider and assessed at the moment of the conformity assessment.\ 

\vspace{0.25cm}

A summary of the AIA's governance mechanism of diachronic identity of high-risk AI systems.\
Requirements specified in Chapter III, Section 2 (Articles 8-15)  define high-level obligations for high-risk AI systems that must be operationalized in practice considering the context, i.e., intended purpose, technical properties as well as deployment environment, of the AI system.\ This occurs through the procedures and documentation that constitute the conformity assessment.\ After deployment, the post-market monitoring system ``continuously'' assesses whether the system still meets the requirements operationalized in the conformity assessment, scanning for substantial modifications, which would require the management of a fresh CE certification for the ``new'' AI system instead.\ We summarize these regulatory components in Figure~\ref{fig:AIA_reg}.\

\section{Claim 1: AIA's Lifecycle Governance Implements \texttt{function}\textsuperscript{\tiny +}'s Diachronic Identity}
\label{section:claim_1}
\begin{figure}[t]
\centering

\resizebox{\linewidth}{!}{%
\begin{tikzpicture}[
  >=Latex,
  cell/.style={
    draw, rounded corners,
    align=left,
    inner sep=6pt,
    text width=4.6cm,
    minimum height=3cm
  },
  head/.style={
    draw, rounded corners,
    align=center,
    inner sep=6pt,
    text width=4.6cm,
    font=\bfseries
  },
  link/.style={Latex-Latex, thick}
]

\def\Xone{0cm}
\def\Xtwo{5.8cm}
\def\Xthree{11.6cm}

\def\Yhead{0cm}
\def\Yone{-3cm}
\def\Ytwo{-7cm}
\def\Ythr{-11cm}

\node[head] (hA) at (\Xone,\Yhead) {EU AIA\\(AI Governance)};
\node[head] (hQ) at (\Xtwo,\Yhead) {AI Identity Questions\\(Governance to Metaphysics)};
\node[head] (hF) at (\Xthree,\Yhead) {\texttt{function}\textsuperscript{\tiny +}\\(AI Metaphysics)};

\node[cell] (A1) at (\Xone,\Yone) {\textbf{Ex-ante conformity assessment}\\
Art.\ 3(20); Arts.\ 8-15; Annex IV.\\
\emph{Requirements:} intended purpose; technical documentation; risk mgmt. and data governance; event recording; transparency around stakeholders and processes.};

\node[cell] (Q1) at (\Xtwo,\Yone) {\textbf{What is the system?}\\
Which techno-function is being assessed, and under which intended purpose, and what counts as appropriate functioning for that context?};

\node[cell] (F1) at (\Xthree,\Yone) {1. \textbf{Techno-function} (AI kind)\\
2. \textbf{Trustworthiness profile}\\
(criteria of appropriate functioning)};

\node[cell] (A2) at (\Xone,\Ytwo) {\textbf{Post-market monitoring}\\
Art.\ 72; Annex IV(9).\\
\emph{Requirements:}  Post-market monitoring plan and system to  collect, document, and analyze data and assess system performance.};

\node[cell] (Q2) at (\Xtwo,\Ytwo) {\textbf{Is the system still compliant over time?}\\
Which monitored signals indicate degradation of appropriate functioning?};

\node[cell] (F2) at (\Xthree,\Ytwo) {\textbf{Trustworthiness level}\\
(time-indexed level of satisfaction of trustworthiness profile)\\
e.g., plateau transitions on a step-wise function.};

\node[cell] (A3) at (\Xone,\Ythr) {\textbf{``Substantial modification''}\\
Art.\ 3(23); Recital 128.\\
\emph{Key decision point:} under a substantial modification, an AI system should be considered to be a new AI system and undergo a new conformity assessment};

\node[cell] (Q3) at (\Xtwo,\Ythr) {\textbf{Same vs. new system?}\\
Was change foreseen? Did it affect compliance or the intended purpose of the system?};

\node[cell] (F3) at (\Xthree,\Ythr) {\textbf{Diachronic identity criterion}\\
Criterion $\textbf{DI}_{\texttt{function}\textsuperscript{\tiny +}}$: same techno-function + same trustworthiness profile + same trustworthiness levels prae-post change.};

\draw[link] (A1.east) -- (Q1.west);
\draw[link] (Q1.east) -- (F1.west);

\draw[link] (A2.east) -- (Q2.west);
\draw[link] (Q2.east) -- (F2.west);

\draw[link] (A3.east) -- (Q3.west);
\draw[link] (Q3.east) -- (F3.west);

\draw[line width=2.5pt] (A1.south) -- (A2.north);
\draw[line width=2.5pt] (A2.south) -- (A3.north);

\draw[line width=2.5pt] (Q1.south) -- (Q2.north);
\draw[line width=2.5pt] (Q2.south) -- (Q3.north);

\draw[line width=2.5pt] (F1.south) -- (F2.north);
\draw[line width=2.5pt] (F2.south) -- (F3.north);
\end{tikzpicture}%
}
\caption{Side-by-side alignment of AIA governance requirements for the lifecycle of high-risk AI systems with the \texttt{function}\textsuperscript{\tiny +} framework.}
\label{fig:AIA_reg}
\end{figure}
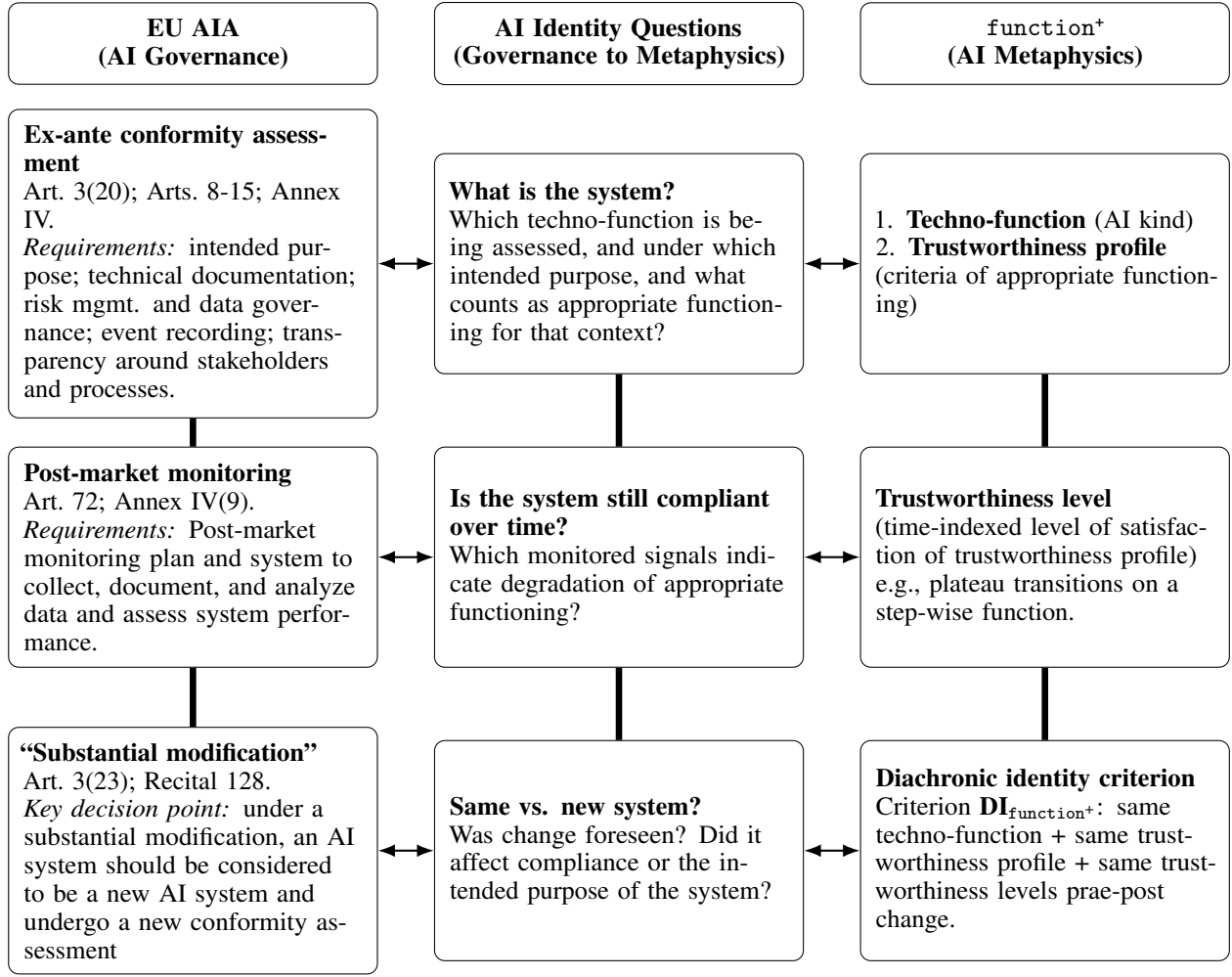

At this stage, it should be clear that the AIA's lifecycle regime for high-risk AI systems operationalizes the diachronic component of the \texttt{function}\textsuperscript{\tiny +} account of artifactual identity. Figure~\ref{fig:AIA_reg} summarizes this alignment: AIA lifecycle obligations are linked to governance-to-metaphysics identity questions, and to the (diachronic) \texttt{function}\textsuperscript{\tiny +} account. 
To see this in some detail, we first note that both approaches require a function-related specification of the AI system, but they do so at different levels. In \texttt{function}\textsuperscript{\tiny +}, the relevant function is the \emph{techno-function}: the core technical operation that identifies the AI kind to which the system belongs. In the AIA, by contrast, the relevant legal notion is \emph{intended purpose}, defined as the use for which the system is intended by the provider, including its specific context and conditions of use. Techno-function and intended purpose should therefore not be simply equated. Rather, intended purpose presupposes or contains a techno-function, but contextualizes it through a use setting, domain, target population, operational constraints, and conditions of deployment. For task-specific high-risk AI systems, the distance between techno-function and intended purpose may be small. For generative and multi-purpose systems, such as large language models, the distance becomes substantial: the same techno-function, such as next-token prediction or autocompletion, may support many intended uses, including drafting, summarization, tutoring, translation, or entertainment. The \texttt{function}\textsuperscript{\tiny +} framework therefore uses techno-function to anchor identity at the level of AI kind, while trustworthiness profiles (and levels) capture the contextualized conditions under which that function counts as appropriate in a given intended use.
We return on this point later in Section \ref{section:recommendations}, when we will collect a few recommendations for the implementation of \texttt{function}\textsuperscript{\tiny +}.\ 
Second, both approaches make use of AI trustworthiness, in the form of contracts under \texttt{function}\textsuperscript{\tiny +} and conformity assessment requirements in the AIA, as a cornerstone of their approaches to diachronic identity.\ We believe there is no bijective mapping between contracts and conformity assessment requirements, where the latter are enumerated in Articles 8-15 in the AIA; first and foremost, it would be difficult to prove the contrary, given the absence of a widespread, standardized template for contracts, which at the time of writing remain a rather high-level construct.\ It seems, however, that, while some conformity requirements are more precise in terms of contracts---such as Article 10 that focuses on fairness---others may encompass distinct contracts---such as Article 13 that discusses explainability, accuracy/performance and human oversight altogether.\
To give shape to this relation, and improve the pragmatics of trustworthiness contracts, practitioners could use, for instance, the AI HLEG's ALTAI and craft their own.\ The conceptual compatibility in the treatment of AI trustworthiness between \texttt{function}\textsuperscript{\tiny +} and the AIA's governance mechanism descends from the fact that contracts collect and discuss the AI requirements that, in the AIA, constitute the conformity assessment of the system.\ However, while \texttt{function}\textsuperscript{\tiny +} provides a necessary and sufficient condition for persistence via the $\textbf{DI}_{\texttt{function}\textsuperscript{\tiny +}}$ criterion, the AIA specifies a sufficient condition only, namely, the emergence of substantial modifications.\ Arguably, this difference stems from the different aims of the metaphysical framework and the AIA.\ In fact, the latter aims to provide a mechanism to recognize substantial modifications and trigger a re-certification procedure.\ For this legal purpose, a sufficient discontinuity condition is enough to trigger renewed assessment, even if it does not amount to a complete metaphysical criterion of persistence. The added value of \texttt{function}\textsuperscript{\tiny +} is precisely to make explicit what is being tracked. The individual dimensions of a trustworthiness profile---for instance accuracy, robustness, fairness, explainability, or cybersecurity---may vary over time as they are measured through post-market monitoring. Such variation does not by itself settle the identity question. What matters for persistence is the trustworthiness level function, namely, the aggregated perspective on whether the system still satisfies its profile to the required degree. This function must therefore be chosen and tested as part of AI governance: it should be robust to reasonable classes of multidimensional variation in the profile, otherwise AI systems would implausibly flicker in and out of existence under ordinary monitoring fluctuations, as noted by \citet{ferrario2025trustworthiness}. At the same time, the function must remain sensitive to changes that are identity-relevant. Persistence is preserved only if the system maintains the same techno-function, the same trustworthiness profile, and the same trustworthiness level. If the aggregated trustworthiness level changes, then the system no longer satisfies the diachronic identity criterion. In AIA terms, this marks the kind of identity-relevant change that qualifies as a substantial modification when it affects compliance with the Chapter III, Section 2 requirements or modifies the intended purpose, thereby triggering renewed conformity assessment. Furthermore, by definition of substantial modification in Recital 128 in the AIA, this condition encodes sameness of techno-function and trustworthiness profile of the diachronic identity criterion $\textbf{DI}_{\texttt{function}\textsuperscript{\tiny +}}$.\ However, at first glance, the trustworthiness level identity that appears in $\textbf{DI}_{\texttt{function}\textsuperscript{\tiny +}}$ is missing in the AIA formulation.\ Upon a closer look, reminders to the quantification of conformity assessment requirements or, in the jargon of \texttt{function}\textsuperscript{\tiny +}, trustworthiness contracts are ubiquitous in the AIA, see, for instance, Recitals 64, 74, 102 and Articles 13 and 15.\ Further, the regulation introduces the expression ``high levels of trustworthiness'' in relation to the mitigation of risks from high-risk AI systems, see Recitals 64 and 123.\ However, what is missing is the precise  quantification of these ``(high) levels'' as a function of the quantifications of its trustworthiness components.\ This is what \texttt{function}\textsuperscript{\tiny +} explicitly provides. We believe this quantification offered by \texttt{function}\textsuperscript{\tiny +} will become especially relevant in the future, given the upcoming specifications for the post-market monitoring plan expected in 2026.\ We will discuss this point in Section \ref{section:recommendations}.

\section{Claim 2: \texttt{function}\textsuperscript{\tiny +} Fills the AIA’s Synchronic Identity Gap}
\label{section:claim_2}
While the AIA regulates diachronic identity extensively by requiring providers to assess whether updates or modifications constitute a ``substantial modification'' and thus, legally, a new system, by contrast, it leaves synchronic identity to Union harmonization legislation---see Annex I in the AIA.\ In fact, the AIA contains no explicit criterion for when two high-risk AI systems (either as safety components of a product, as products themselves, or as stand-alone systems) placed on the market should be considered legally ``the same'' system at any time.\ Logically, this division of labor avoids duplication of efforts, as mentioned in Recital 124 and Articles 8 and 72 in the AIA: existing product regulations may already contain implicit or explicit synchronic identity criteria.\ However, this approach is not free from limitations that deserve our attention.\ 

First, no criteria for synchronic identity of stand-alone high-risk AI systems are provided. (This applies to generative AI systems as well.)\ In fact, by definition, no existing Union harmonization legislation can apply to these systems.\ As a result---see the synchronic identity use case in Section \ref{section:motivating_examples}---this gap implies that providers and governance bodies lack a basis for determining whether two stand-alone high-risk AI systems should be regarded as identical.\ Second, for high-risk AI systems that are safety components of products and services, it is not clear how the relevant Union harmonization legislation is to be applied.\ For instance, in EU pharmaceutical law, a medicinal product applicant is not required to document the results of toxicological and pharmacological tests or the results of clinical trials if they can prove that the medicinal product is ``essentially similar'' to a medicinal product authorized in the Member State and has the same intended ``therapeutic use'' among others---see Article 10 in \citep{Directive2001_83_EC}.\ In the EU Medical Device Regulation (MDR), the notion of ``equivalence'' permits clinical trial data from one device to be used in the clinical evaluation of another device, provided that technical, biological, and clinical similarities between the two devices can be demonstrated---see Annex XIV, PART A, (3) in \citep{EU2017MDR}.\footnote{Note that the term ``equivalence'' does not appear in the AIA.\ The adjective ``equivalent'' does, but it is not referred to synchronic identity---e.g., it is used in the context of general AI systems with systemic risk, see Article 51 in the AIA.}  Ultimately, the MDR requires that claims of device equivalence be supported by sound scientific reasoning, and that manufacturers have adequate access to the reference device's data to substantiate such claims.\ 
The problem is that identity criteria in this and other regulations target products as a whole, not AI subsystems embedded as safety components.\ This generates a mereological problem: since Baker's seminal work on the ontology of artifacts \citep{baker2004ontology}, modern philosophy treats constitution and identity as distinct relations, and the criteria for establishing the identity of a component need not coincide with those of the whole they refer to \citep{gallois2016metaphysics}.\ (For instance, identity criteria of water molecules are different from identity criteria of rivers they are part of.) Further, even where regulations such as the MDR provide proxies for synchronic identity under the notion of equivalence, these criteria rely heavily on expressions such as ``(essentially) similar'', and ``same'' with the effect of diluting definitional precision.\ Equivalence across technical, biological, and clinical dimensions may be treated as a proxy for trustworthiness, but it remains an underspecified criterion for AI synchronic identity, especially when combined with the mereological challenge mentioned above. (Think about the ubiquitous use of open-source generative AI models in products and services, such as AI agents.)

\begin{figure}[t]
\centering

\resizebox{\linewidth}{!}{%
\begin{tikzpicture}[
  >=Latex,
  sys/.style={draw, rounded corners, align=center, inner sep=6pt, text width=4.2cm, minimum height=1.6cm},
  big/.style={draw, rounded corners, align=left, inner sep=8pt, text width=5cm, minimum height=3.8cm},
  link/.style={->, thick}
]

\def\Xsys{0cm}
\def\Xassess{6.2cm}
\def\Xpms{13.0cm}
\def\Xout{19.8cm}

\def\Yno{1.1cm}
\def\Yyes{-1.1cm}

\def\Ymid{0cm}

\node[sys] (A) at (\Xsys,\Yno) {\textbf{System} $A$};
\node[sys] (B) at (\Xsys,\Yyes) {\textbf{System} $B$};

\node[big] (CA) at (\Xassess,\Ymid) {\textbf{Conformity Assessment}\\[2mm]
\emph{Question:} do the systems\\
(1) have the same intended purpose, and\\
(2) the same trustworthiness profile?};

\node[big] (PMS) at (\Xpms,\Ymid) {\textbf{Post-market Monitoring}\\[2mm]
\emph{Question:} at time $t$, do the systems\\
have the same trustworthiness level?};

\node[sys] (YES2) at (\Xout,\Yyes) {The systems are the same system for regulatory purposes};

\node[sys] (NO)   at (\Xout,\Yno)  {The systems \textbf{are not} the same system for regulatory purposes};

\draw[link] (A.east) -- (CA.west);
\draw[link] (B.east) -- (CA.west);

\draw[link] (CA.east) -- node[above]{yes} (PMS.west);

\draw[link] (PMS.east) -- node[above]{yes} (YES2.west);

\draw[link] (PMS.east) -- node[above]{no} (NO.west);

\draw[->, thick]
  (CA.north) -- ++(0,1cm)
  -- node[midway, above, xshift=4mm]{no} ++(13.6cm,0)
  -- (NO.north);

\draw[->, thick]
  (PMS.north) -- ++(0,0.4cm)
  -- node[midway, above, xshift=4mm]{no} ++(6cm,0)
  -- (NO.north);

\end{tikzpicture}%
}

\caption{\textbf{Claim 2}: Synchronic identity decision flow. Systems $A$ and $B$ are the ``same'' for regulatory purposes only if they have the same intended purpose and trustworthiness profile at conformity assessment, and their trustworthiness level matches at time $t$ under post-market monitoring.}
\label{fig:claim2-flow}
\end{figure}
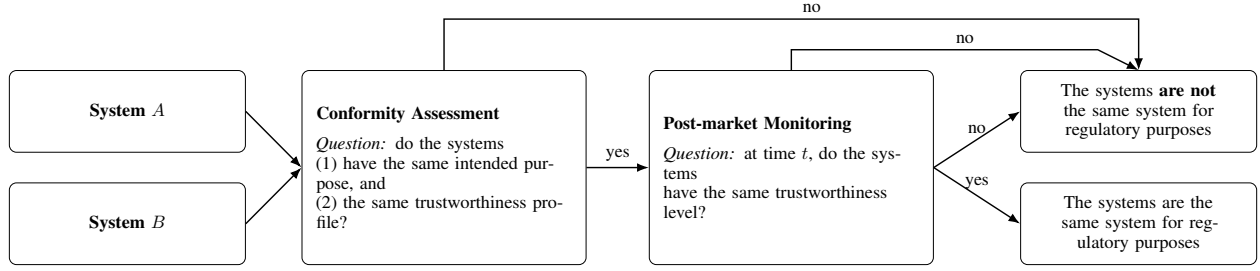

In summary, synchronic identity matters whenever governance must decide whether compliance evidence, documentation, monitoring signals, and corrective actions for one high-risk AI system instantiation legitimately generalize to another instantiation at a given point in time, e.g., across certification scoping, traceability issues, incident response cases, comparability issues with white-labeling, and procurement.
While this is something other EU regimes handle via grouping identifiers and ``essential characteristics,'' stand-alone high-risk AI lacks an equally auditable sameness criterion under the AIA.
Three recurring governance situations make the AIA's synchronic identity gap practically salient. \textbf{Market surveillance and corrective-action propagation:} when a serious incident, complaint, or non-compliance finding concerns one stand-alone high-risk AI instantiation, authorities must decide whether mitigation, rollback, user notification, or withdrawal obligations generalize to other contemporaneous instantiations marketed as ``different'' systems. \textbf{White-labeling and reseller variants:} providers may distribute functionally comparable stand-alone high-risk systems under different product names, interfaces, or integration wrappers; without an auditable sameness criterion, documentation and monitoring evidence can be fragmented across labels, undermining traceability and comparability. \textbf{Procurement and equivalence disputes:} contracting authorities must decide whether two bids are substantively comparable---for instance, whether a premium-priced variant is materially distinct or whether conformity evidence and trustworthiness claims for one variant can be relied upon for another. (This is the case we explored in Section~\ref{section:motivating_examples}.)
Figure~\ref{fig:claim2-flow} makes this governance problem explicit: it represents a minimal audit workflow in which sameness claims must be justified first at conformity assessment (purpose and profile) and then maintained under post-market monitoring (trustworthiness level at time $t$). The figure also highlights when a regulatory non-sameness outcome obtains, making the synchronic criterion operationally legible. Appendix~\ref{app:audit} (Table~\ref{tab:audit_minimal}) translates this decision flow into a minimal set of auditable fields and indicates where each field should appear in AIA lifecycle artifacts.
Concretely, \(\textbf{SI}_{\texttt{function}\textsuperscript{\tiny +}}\) turns a  sameness claim into a three-step audit question: do the systems share the same intended purpose, the same trustworthiness profile, and the same trustworthiness level at the relevant time? If any of these elements fails, the systems should not be treated as the same for regulatory comparability---see Figure~\ref{fig:claim2-flow}.
In summary, the \texttt{function}\textsuperscript{\tiny +} approach to synchronic identity and, in particular, criterion $\textbf{SI}_{\texttt{function}\textsuperscript{\tiny +}}$, provide the definitional precision needed to close this gap in the AIA, anchoring the comparison of AI systems in techno-function, intended purpose, and requirements of appropriate functioning.\ However, its utilization is conditional on the implementation of a few recommendations we discuss in the following section.

\section{Recommendations for Regulating AI System Identity}
\label{section:recommendations}
Through the use cases in Section \ref{section:motivating_examples}, we showed that clarifying AI identity is key for effective governance. In Sections \ref{section:claim_1} and \ref{section:claim_2} we then argued that \texttt{function}\textsuperscript{\tiny +} aligns with the AIA's lifecycle treatment of diachronic identity and supplies a feasible approach for synchronic identity. However, its practical uptake is limited by underspecified components, especially the precise statement of techno-function and intended purpose and the operationalization of measurable trustworthiness profiles. Let us address these gaps, providing a few recommendations to improve the applicability of \texttt{function}\textsuperscript{\tiny +} in the AI regulation domain. We refer to Appendix~\ref{app:audit}, Table~\ref{tab:audit_minimal}, for a minimal checklist that maps our recommendations to specific AIA artifacts (intended purpose statement, risk management file, and post-market monitoring plan) and to the decision points in Figure~\ref{fig:claim2-flow}.

\subsubsection{Test and document high-risk AI system functions in relation to their intended purpose.}
What is the function of a large language model-based conversational agent such as ChatGPT? In the sense relevant to \texttt{function}\textsuperscript{\tiny +}, the answer concerns its \emph{techno-function}: for paradigmatic large language models, this is next-token prediction or autocompletion, operationalized as the generation of text continuations from prompts. This should be distinguished from the system's \emph{intended purpose} in the AIA sense, namely, the use for which the system is provided, including its specific context and conditions of use. The same techno-function can support many intended uses: creating poetry, summarizing documents, drafting code, assisting education, or producing administrative text. These uses may differ in their social meaning, risk profile, user expectations, and regulatory relevance, even when the underlying techno-function remains the same.
This distinction matters for applying \texttt{function}\textsuperscript{\tiny +}. The techno-function anchors the identity of the AI kind; the intended purpose contextualizes that function for governance. Philosophy of technology helps articulate this distinction by separating \emph{techno-functions} (core technical operations), \emph{use-functions} (practical affordances for users), and \emph{social functions} (societal roles emerging from interaction with the system) \citep{vermaas2003ascribing}. For instance, a conversational agent may have the techno-function of next-token autocompletion while being used as a summarizer, a tutor, a therapist-like companion, or a ``stern reviewer of scientific papers.'' These social functions affect the system's societal impact and the vulnerability of its users and need to be considered in the regulatory mechanism of the system's identity: auditing of generative AI technology is a timely and complex challenge  \citep{mokander2025blueprint,mokander2024auditing}.
Thus, specifying the AI techno-function is necessary for anchoring system identity over time, but it is not sufficient for regulating the effects of its deployment. Providers should therefore document both the techno-function and the intended uses for which the system is offered, including context and conditions of use, explicit exclusions, non-intended uses, and non-compliant uses. Governance bodies should encourage or require this documentation in conformity assessment, post-market monitoring, and the assessment of substantial modifications.
Further, non-compliant intended uses should be enumerated and discussed as much as possible.\ A prominent class of such non-compliant uses is promoted by the rise of AI anthropomorphization \citep{floridi2024anthropomorphising,shanahan2023role,ferrario2025social,akbulut2024all,ferrario2026anthrop}. Pre-market, this points to extensive social role-sensitive testing; post-market, to guardrails that surface limits to users and reduce ``social misattributions'' due to multi-purpose conversational and agentic affordances.

\subsubsection{Promote standardization of contracts, conformity, post-market assessments, and trustworthiness reporting.}
In the absence of standardized reporting of trustworthiness contracts, assessments of AI identity may vary across EU countries, and market surveillance authorities may struggle to determine whether functionally identical systems are truly distinct or whether equivalent systems receive divergent regulatory treatment.\ Although it is not the goal of a regulation such as the AIA to provide exhaustive details on trustworthiness contracts, conformity, and post-market assessments,\footnote{Authors have already raised criticism against the material scope and other definitional ambiguities of the AIA---see, for instance, \cite{mokander2022conformity,nolte2025robustness,laux2024trustworthy}.\ Furthermore, the requirements specified in Chapter III, Section 2 (Articles 8-15) in the AIA define context-independent obligations for high-risk AI systems.\ Finally, the contextualization of these obligations, e.g., an employment opportunity domain vs. healthcare, should occur through the procedures and documentation that constitute the conformity assessment and the guardrails in the post-market monitoring system and would further increase complexity.
} the vagueness in interpretation of the levels of detail in these procedures may encourage exploiting ambiguity to avoid re-assessment among organizations competing in the EU market.\
Providers may try to avoid excessive procedural burden and triggering of the conformity re-assessment disproportionately, a fact that would have a serious impact on their system operations, incurring costs and important investment of resources to obtain a fresh conformity certification.\ 
However, the idea that all permissible changes of a high-risk AI system's components could be anticipated, analyzed for the conformity assessment and assessed during post-market monitoring is practically untenable.\ 
These challenges arise especially with generative AI-based systems, where intended use is vague, changes are hard to anticipate, and the documentation burden may easily become prohibitive.\


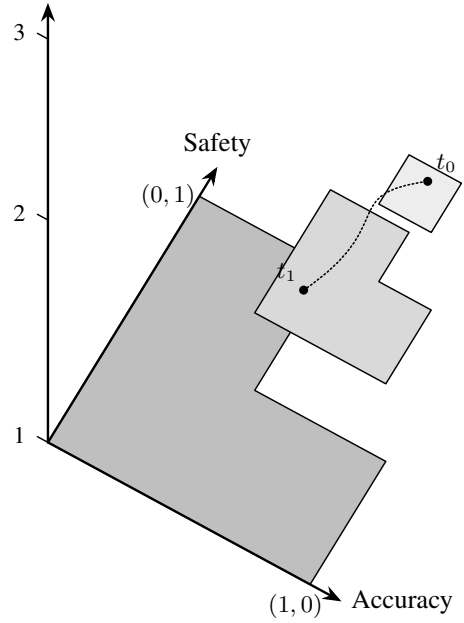
\begin{figure}[t]
\centering
\tdplotsetmaincoords{20}{30} 
\begin{tikzpicture}[tdplot_main_coords, scale=4, line join=round, line cap=round]

\tikzset{
  traj/.style={line width=0.6pt, densely dotted},
  pt/.style={circle, fill, inner sep=1.2pt},
  lab/.style={font=\small},
  axis/.style={-{Stealth[length=2.5mm]}, line width=0.9pt},
}

\def\zOne{1}
\def\zTwo{1.75}
\def\zThr{2.5}

\coordinate (L1A) at (0,0,\zOne);
\coordinate (L1B) at (1,0,\zOne);
\coordinate (L1C) at (1,0.5,\zOne);
\coordinate (L1D) at (0.5,0.5,\zOne);
\coordinate (L1E) at (0.5,1,\zOne);
\coordinate (L1F) at (0,1,\zOne);

\coordinate (L2A) at (0.5,1,\zTwo);
\coordinate (L2B) at (0.5,0.5,\zTwo);
\coordinate (L2C) at (1,0.5,\zTwo);
\coordinate (L2D) at (1,0.8,\zTwo);
\coordinate (L2E) at (0.8,0.8,\zTwo);
\coordinate (L2F) at (0.8,1,\zTwo);

\coordinate (L3A) at (0.8,0.8,\zThr);
\coordinate (L3B) at (0.8,1,\zThr);
\coordinate (L3C) at (1,1,\zThr);
\coordinate (L3D) at (1,0.8,\zThr);

\filldraw[draw, line width=0.55pt, fill=black!25] (L1A)--(L1B)--(L1C)--(L1D)--(L1E)--(L1F)--cycle;
\filldraw[draw, line width=0.55pt, fill=black!15] (L2A)--(L2B)--(L2C)--(L2D)--(L2E)--(L2F)--cycle;
\filldraw[draw, line width=0.55pt, fill=black!7 ] (L3A)--(L3B)--(L3C)--(L3D)--cycle;

\coordinate (tOne) at (0.60,0.65,\zTwo);
\coordinate (tZero) at (0.90,0.95,\zThr);

\node[pt] at (tOne)  {};
\node[lab, anchor=south east, xshift=1pt] at (tOne) {$t_1$};

\node[pt] at (tZero) {};
\node[lab, anchor=south west, xshift=-0.2pt] at (tZero) {$t_0$};

\draw[traj]
  (tOne) .. controls (0.85,0.85,2.25) and (0.7,0.7,2.70) .. (tZero);

\draw[axis] (0,0,\zOne) -- (1.12,0,\zOne) node[anchor=west] {Accuracy};
\draw[axis] (0,0,\zOne) -- (0,1.12,\zOne) node[anchor=south] {Safety};
\draw[axis] (0,0,\zOne) -- (0,0,5.25) node[anchor=west] {};

\node[lab, anchor=north, xshift=-5.5pt] at (1,0,\zOne) {$(1,0)$};
\node[lab, anchor=west, xshift=-25.5pt]  at (0,1,\zOne) {$(0,1)$};

\foreach \zz/\lbl in {\zOne/1,\zTwo+1.4/2,\zThr+2.4/3}{
  \draw[line width=0.55pt] (0,0,\zz) -- (-0.04,0,\zz);
  \node[lab, anchor=east] at (-0.05,0,\zz) {\lbl};
}

\end{tikzpicture}
\caption{Illustrative piecewise-constant trustworthiness levels over accuracy and safety. The system exhibits three plateaus (high, medium, and low levels of trustworthiness). Small fluctuations within a plateau do not change the overall system trustworthiness, as shown by the portion of the accuracy-safety curve between $t_0$ and $t_1$.  Conformity re-assessment should be considered when a plateau boundary is crossed, e.g., a transition from level three (high) to two (medium) occurs at $t_1$.}
\label{fig:trustworthiness}
\end{figure}

Thus, regulators should promote standardization of reporting practices for trustworthiness contracts, conformity, and post-market monitoring, as in medical device regulation (e.g., structured post-market clinical follow-up templates in the MDR) and cybersecurity certification (e.g., NIST SP 800-53 Rev.\ 5 control families).\ 
Developing analogous templates and taxonomies for AI would require time, expert working groups, and iterative revision, but the earlier such processes start, the more feasible they become.\ From the point of view of standardization, however, reliance on trustworthiness contracts does not lead to an increase in complexity as they respond to AIA's conformity assessment requirements---see Section \ref{section:claim_1}.\ Existing initiatives already point in this direction: ISO/IEC JTC 1/SC 42 has issued standards on AI lifecycle management\footnote{\url{https://www.iso.org/committee/6794475.html}} and CEN/CENELEC JTC 21 is developing harmonized standards in support of the AI Act.\footnote{\url{https://jtc21.eu/}}
However, we argue that regulating AI identity based on contracts individually and the changes affecting them is both inefficient and prone to over-sensitivity: small deviations in any of the dimensions constituting AI trustworthiness may trigger unnecessary regulatory alerts.\ What matters is not each component in isolation but the system's overall capacity to remain worthy of trust at a given point in time, as suggested by \texttt{function}\textsuperscript{\tiny +}'s treatment of AI trustworthiness levels---see $\textbf{DI}_{\texttt{function}\textsuperscript{\tiny +}}$ and $\textbf{SI}_{\texttt{function}\textsuperscript{\tiny +}}$.\ In other regulated domains, this challenge is addressed through composite or global assessments: cars receive `roadworthiness' certificates (e.g., Euro NCAP star ratings) that condense numerous safety checks into a single judgment; similar considerations hold for other machinery, such as aircraft, as well as medical devices (e.g., ISO 14971 risk indices integrating performance, safety, and usability).\ By analogy, the standardization of trustworthiness reporting for AI should move toward specifying an aggregated perspective, namely a trustworthiness level function  integrating dimensions such as accuracy, safety, robustness, fairness, and explainability into a composite indicator.\  Thus, a temporary decline in one dimension could be tolerable if others remain high, but sustained or joint reduction would indicate a degradation in trustworthiness and an end to the AI system's persistence and the triggering of a new conformity assessment. A practical, but non-exclusive, way to model AI trustworthiness levels is with multi-dimensional piecewise-constant functions that map regions of the metric space of trustworthiness-relevant dimensions (e.g., accuracy, safety, and transparency) to a finite number of trustworthiness plateaus. These simple and interpretable functions can be learned by classification and regression trees, providing a data-driven approach to the standardization of AI trustworthiness reporting. Figure~\ref{fig:trustworthiness} provides an illustration. Here, what matters is that the chosen trustworthiness level function, including its dimensions, thresholds, aggregation rules, and robustness to reasonable multidimensional variation, is explicitly justified and auditable. On the one hand, as part of the conformity assessment, providers should disclose how the trustworthiness-level function is derived, including a detailed description of training data.\footnote{Alternatives are possible: rule-based specifications or a scoring scheme.\ Similar considerations on reporting apply.} 
Weightings, metric choices, and aggregation rules for trustworthiness-level functions are not assumed to be universal: they are context- and designer-dependent, reflecting domain-specific risk priorities, stakeholder expectations, and operational constraints. \citet{rabanser2026towards} illustrate this approach in the context of AI agents by identifying relevant trustworthiness dimensions, e.g., robustness and safety, operationalizing them via concrete measures, aggregating them into a composite assessment, and employing the resulting scoring scheme in empirical studies that include established benchmarks.
On the other hand, as part of standardization efforts, governance bodies can adopt such metrics as part of standardized post-market monitoring, reducing opportunities for providers to exploit loosely specified regulations to avoid new conformity assessments and making substantial modifications auditable in practice.

 What matters for governance is that these choices are \emph{explicitly declared and justified} in a standardized, comparable format, so that oversight bodies can audit trade-offs consistently across providers and track them over time.

\section{Conclusions}
\label{section:conclusions}
The AIA is the most ambitious attempt so far to regulate AI in Europe.\ By requiring conformity assessments, substantial modification checks, and post-market monitoring, it implicitly commits to a position on the diachronic identity of AI systems. Yet, it leaves synchronic identity unaddressed, except by reference to sectoral legislation.\ This asymmetry risks creating regulatory blind spots in procurement, liability, and market surveillance.\ We argued that the \texttt{function}\textsuperscript{\tiny +} framework can fill this gap, providing both identity criteria within a single frame, clarifying intended purpose, trustworthiness commitments, and measurable indicators for conformity and monitoring.\ As a result, it provides a metaphysical framework for the AIA’s lifecycle approach while supplying the missing criteria for synchronic cases.\ 
The philosophical analysis proposed in this work is a conceptual and auditing lens that can help make identity-relevant determinations in high-risk AI system lifecycle documentation transparent and contestable within the AIA framework. It does not replace legal interpretation, technical standards, conformity assessment procedures, or empirical validation: it helps make identity-relevant determinations in high-risk AI system lifecycle documentation more explicit, transparent, and contestable within the AIA framework.
Clarifying identity conditions of AI systems helps make legal obligations enforceable, procurement fair, and accountability tractable across traceability and comparability issues, and incident responses. As generative and agentic AI proliferate across domains, metaphysical clarity will be needed for the governance of the technologies that increasingly shape our societies.

\section*{Acknowledgments}
The author acknowledges partial support by the Swiss National Science Foundation (SNSF), grant no. 229061.
The author thanks Michele Loi for insightful comments on a preliminary version of this work.

\newpage
\section*{Appendix}
\subsection*{Minimal audit checklist for synchronic identity}
\label{app:audit}
Table~\ref{tab:audit_minimal} displays minimal auditable content for synchronic identity (\textbf{Claim 2}) and its placement in AIA artifacts. The table operationalizes Figure~\ref{fig:claim2-flow} as a checklist for audit, procurement, and dispute contexts.

\begin{table}[h!]
\small
\centering
\caption{Minimal auditable content for synchronic identity (Claim 2) and its placement in AIA artifacts. The table operationalizes Figure~\ref{fig:claim2-flow} as a checklist for audit, procurement, and dispute contexts.}
\label{tab:audit_minimal}
\begin{tabularx}{\linewidth}{@{}p{2.6cm}p{3.6cm}X@{}}
\toprule
\textbf{Decision point} & \textbf{AIA artifact (where)} & \textbf{Minimal auditable fields (what)} \\
\midrule

\textbf{D1: Intended purpose match} &
Annex IV technical documentation; instructions for use &
Declared intended purpose; context and conditions of use; target population; deployment constraints; explicit exclusions / non-intended uses; stated dependencies on environment assumptions (e.g., sensors, latency constraints, human oversight conditions). \\

\addlinespace
\textbf{D2: Trustworthiness profile match} &
Annex IV (metrics and justification); risk management documentation &
Trustworthiness dimensions claimed (e.g., robustness, safety, fairness, transparency); operational definitions and metrics; evaluation protocols and datasets; acceptance thresholds; aggregation rule (incl.\ weightings) with justification; known limitations and trade-offs. \\

\addlinespace
\textbf{D3: Trustworthiness level match at time $t$} &
Post-market monitoring plan and records; change logs &
Current measured levels at time $t$ (per D2 metrics); monitoring cadence; drift/degradation indicators and triggers; incident history relevant to the profile; version/configuration identifiers; evidence that the level uses the same aggregation rule as in D2. \\

\addlinespace
\textbf{Outcome: ``same'' for regulatory comparability} &
Audit / procurement / surveillance file &
Explicit rationale that D1--D3 are satisfied; scope of evidence portability (what can be re-used: tests, documentation, corrective actions); validity window and re-check triggers. \\

\addlinespace
\textbf{Outcome: ``not same'' for regulatory comparability} &
Audit / procurement / surveillance file &
Failure point (D1, D2, or D3) and evidence; implication statement: compliance evidence does not generalize; required next step (separate assessment scope, additional documentation, or separate monitoring). \\
\bottomrule
\end{tabularx}
\end{table}

\newpage

\bibliographystyle{plainnat} 
\bibliography{bibliography} 

\end{document}